\documentclass[aps,prb,twocolumn,amsfonts,floatfix]{revtex4}

\usepackage{graphicx}
\usepackage{amsmath}

\newcommand{\dimm}{D} % define new macro if you wish
\begin{document}

\title{Quantum chaos: An introduction via chains of interacting spins 1/2}

\author{Aviva Gubin}
\author{Lea F. Santos}
\email{lsantos2@yu.edu}
\affiliation{Department of Physics, Yeshiva University, 245 Lexington Avenue, New York, NY 10016, USA}

%\date{\today}

\begin{abstract}
We introduce aspects of quantum chaos by analyzing the eigenvalues and the eigenstates of quantum many-body systems. The properties of quantum systems whose classical counterparts are chaotic differ from those whose classical counterparts are not chaotic. The spectrum of the first exhibits repulsion of the energy levels. This is one of the main signatures of quantum chaos. We show how level repulsion develops in 
one-dimensional systems of interacting spins 1/2 which are devoid of random elements and involve only two-body interactions. In addition to the statistics of the eigenvalues, we analyze how the structure of the eigenstates may indicate chaos. The programs used to obtain the data are available online. 
\end{abstract}

\maketitle

\section{Introduction}

Classical chaos is related to the extreme sensitivity of the dynamics of a system to its
initial conditions, a concept that can be traced back to Poincar\'e.\cite{PoincareBook}
The main features of classical chaos can be illustrated 
by a dynamical billiard, which is an idealized billiard table with no friction
where a particle reflects elastically from
boundaries which can have any shape. The motion of the particle 
is represented in phase space by a trajectory whose evolution 
is restricted to a surface of constant energy.
Depending on the shape of the boundaries, the system may be
chaotic, which means that two trajectories whose initial conditions are very close will diverge exponentially in time. The rate of this separation is characterized 
by the Lyapunov exponent.\cite{GleickBook}
The trajectories may also become ergodic, which implies that after a long time the particle will have visited the entire
surface of constant energy. Equivalently, we may
say that after a long time, the particle is equally likely to be found in any 
point of the accessible phase space.

For quantum systems the notion of phase-space trajectories 
loses its meaning as can be seen from the Heisenberg uncertainty principle. 
Nevertheless, because classical physics is a limit of quantum physics, it is natural to 
search for quantum signatures of classical chaos. The relation between quantum mechanics and classical chaos is the 
subject of quantum chaos.\cite{Jensen1992,Gutzwiller1992,Rudnick2008}

Quantum chaos is a very broad field. Our intention is not to review the 
entire field, but to focus on a particular aspect of it, namely, the 
spectral statistics and structures of eigenstates of 
quantum many-body systems as they transition from the regular to the chaotic domain. 
As a preparation for our goal, we first give a brief idea of some of the important topics of the field
(more details are in Refs.~\onlinecite{GutzwillerBook,HaakeBook,ReichlBook,StockmannBook} and in the references to be cited).

Quantum chaos encompasses
semiclassical theories aiming at establishing direct links between the trajectories of 
classically chaotic systems and the properties of these systems in the quantum domain.
The main advances in this area were achieved by periodic-orbit theory, which 
provides a way to calculate the spectrum of
a quantum system from its classical periodic orbits (trajectories that repeat themselves after a 
certain time). An excellent introduction to the subject
was given by Gutzwiller.\cite{Gutzwiller1992} Gutzwiller's method allowed for the development of a 
theory of ``scars,''\cite{Berry1989,StockmannBook}
which refer to the structure of eigenstates that concentrate
along the classical periodic orbits of chaotic systems. Scars were
systematically studied by Heller,\cite{Heller1984} and the first
experimental observations occurred in quantum billiards. \cite{Jensen1992}
Quantum billiards obey the laws of quantum mechanics and
correspond to miniature versions of dynamical billiards. 

Another important step in the understanding of quantum chaos 
came from the verification
that the distribution of the spacings between neighboring energy levels of a quantum billiard
depend on the billiard's classical 
counterpart.\cite{McDonald1979,Berry1981,Bohigas1984,Rudnick2008} 
If the latter is chaotic,
the energy levels are highly correlated and repel each other; if it
is regular (integrable), the energy levels are uncorrelated, randomly distributed, and can cross.

Level repulsion is one of the main features of quantum chaos.
It has been observed in other quantum systems, such as
atoms in strong magnetic fields \cite{Jensen1992} and
systems of coupled particles,\cite{HaakeBook,ReichlBook,StockmannBook,Guhr1998}.
Interestingly, in the context of number theory, level repulsion has been associated also with distributions of prime numbers.\cite{Timberlake2006}

In this article we focus on 
quantum many-body systems. 
A common approach when dealing with such systems is to ignore the details
of the interactions and treat them statistically with random matrices.
The idea is that when the interactions are strong and 
the behavior of the system is sufficiently complex, generic
properties should emerge. 
This approach was taken by Wigner \cite{Wigner1951} to describe the spectrum of heavy nuclei.
He employed matrices with random elements 
whose only constraint was to satisfy the symmetries of the system.
The level spacing distributions of these matrices showed level 
repulsion and agreed surprisingly well with 
the data from actual nuclei spectra. 
When level repulsion was later verified in billiards,
a connection between quantum chaos and random matrices became established.
Soon after the introduction of random matrices in nuclear physics,
they were employed in 
the analysis of the spectrum of other quantum many-body systems, such as atoms, molecules, and
quantum dots.\cite{HaakeBook,ReichlBook,StockmannBook,Guhr1998,Beenakker1997,Alhassid2000}

The application of random matrix theory is not
restricted to the statistics of eigenvalues, but accommodates also studies of eigenstates.
Eigenstates of random matrices are pseudo-random vectors; that is,
their amplitudes are random variables.\cite{Berry1977,Brody1981} 
All the eigenstates are statistically similar, they spread through all basis vectors with
no preferences and are therefore ergodic.

Despite the success of random matrix theory in describing spectral statistical properties, 
it cannot capture the details of real quantum many-body systems. 
The fact that random matrices are completely filled 
with statistically independent elements implies 
infinite-range interactions and the simultaneous interaction of many particles.
Real systems have few-body (most commonly only two-body) interactions which are usually finite range.
A better picture of systems with finite-range interactions is provided by 
banded random matrices, which were also studied by Wigner. \cite{Wigner1955}
Their off-diagonal elements are random and statistically independent,
but are non-vanishing only up to a fixed distance from the diagonal.
There are also ensembles of random matrices that take into account
the restriction to few body interactions, so that only the elements associated
with those interactions are nonzero; an example is the
two-body-random-ensemble\cite{French1970,BohigasFlores1971,Flores2001} (see 
reviews in Refs.~\onlinecite{Brody1981,Kota2001}). Other
models which describe systems with short-range and few-body interactions do not include random elements, such as
nuclear shell models,\cite{ZelevinskyRep1996} and the
systems of interacting spins which we consider in this article.

All the matrices we have mentioned can lead to level repulsion, but differences are observed.
For instance, eigenstates of random matrices are completely spread (delocalized) in any basis,
whereas the eigenstates of systems with few-body interactions delocalize only
in the middle of the spectrum.\cite{Kota2001,ZelevinskyRep1996,Izrailev1990,Casati1993,Flambaum1994} 

In this paper we study a one-dimensional system of interacting spins 1/2. The system involves only nearest-neighbor interactions, and in some cases, 
also next-nearest-neighbor interactions. 
Depending on the strength of the couplings, the system may develop chaos, which 
is identified by calculating
the level spacing distribution. We also compare 
the level of delocalization of the eigenstates 
in the integrable and chaotic domains. It is significantly larger in the 
latter case, where the most delocalized states are found in the middle of the spectrum.

The paper is organized as follows. Section~II provides a detailed description of the
Hamiltonian of a spin 1/2 chain. Section~III explains how to compute 
the level spacing distribution and how to quantify the level of delocalization of the eigenstates.
Section~IV shows how the mixing of symmetries may erase level repulsion even when the system is chaotic. Final remarks are given in Sec.~V.

\section{Spin-1/2 chain}
\vskip -0.1 cm
We study a one-dimensional spin 1/2 system (a spin 1/2 chain) described by the Hamiltonian 
\begin{subequations}
\label{hamall}
\begin{align}
H &= H_z + H_{\text{NN}},
\label{ham}\\
\noalign{\noindent where}
H_z &= \sum_{i=1}^{L} \omega_i S_i^z = \left( \sum_{i=1}^{L} \omega S_i^z \right) + \epsilon_{d} S_{d}^z \label{ham2}\\
H_{\text{NN}} &= \sum_{i=1}^{L-1} \left[ J_{xy} \left(
S_i^x S_{i+1}^x + S_i^y S_{i+1}^y \right) +
J_z S_i^z S_{i+1}^z \right].\label{ham3}
\end{align}
\end{subequations}
We have set $\hbar$ equal to 1,
$L$ is the number of sites, $S^{x,y,z}_i = \sigma^{x,y,z}_i/2$ 
are the spin operators at site $i$, and  $\sigma^{x,y,z}_i$ are the Pauli matrices. 
The term $H_z$ gives the Zeeman splitting of each spin $i$,
as determined by a static magnetic field in the $z$ direction. All sites are assumed to have the same energy splitting
$\omega$, except a single site $d$, whose energy splitting $\omega+\epsilon_d$ is caused
by a magnetic field slightly larger than the field
applied on the other sites. This site is referred to as a defect.

A spin in the positive $z$ direction (up) is indicated by $|\uparrow\rangle$ or by
the vector $\binom{1}{0}$; a spin in the negative $z$ direction (down) is represented by
$|\downarrow\rangle$ or $\binom{0}{1}$. 
An up spin on site $i$ has energy $+\omega_i/2$, and a down spin has energy $-\omega_i/2$.
A spin up corresponds to an excitation.

The second term, $H_{\text{NN}}$, is known as the XXZ Hamiltonian.
It describes the couplings between nearest-neighbor (NN) spins; $J_{xy}$ is the strength 
of the flip-flop term $S_i^x S_{i+1}^x + S_i^y S_{i+1}^y$,
and $J_z$ is the strength of the Ising interaction $S_i^z S_{i+1}^z$. The flip-flop term
exchanges the position of neighboring up and down spins according to 
\begin{equation}
J_{xy}(S_i^x S_{i+1}^x + S_i^y S_{i+1}^y)|\uparrow_i \downarrow_{i+1} \rangle = 
(J_{xy}/2) |\downarrow_i \uparrow_{i+1} \rangle,
\label{flipflop}
\end{equation}
or, equivalently, it moves the excitations through the chain. We have assumed 
open boundary conditions as indicated by the sum 
in $H_{\text{NN}}$ which goes from $i=1$ to $L-1$. Hence, an excitation in site 1 (or $L$) 
can move only to site 2 (or to site $L-1$). Closed boundary conditions, where an excitation in site 1 can move also to site $L$ (and vice-versa) are mentioned briefly in Sec.~IV.

The Ising interaction implies that pairs of parallel spins have higher energy than pairs of anti-parallel
spins, that is,
\begin{equation}
J_z S_i^z S_{i+1}^z |\uparrow_i \uparrow_{i+1}\rangle =+(J_z/4) |\uparrow_i \uparrow_{i+1}\rangle,
\label{upup}
\end{equation}
and
\begin{equation}
J_z S_i^z S_{i+1}^z |\uparrow_i \downarrow_{i+1}\rangle =-(J_z/4) |\uparrow_i \downarrow_{i+1}\rangle.
\label{updown}
\end{equation}

For the chain described by Eqs.~(\ref{hamall}) the total 
spin in the $z$ direction, $S^z=\sum_{i=1}^L S_i^z$, is conserved, that is,
$[H,S^z]=0$. This condition means that the total number of
excitations is fixed; the Hamiltonian cannot create or annihilate excitations, it 
can only move them through the chain.

To write the Hamiltonian in matrix form and diagonalize it to find
its eigenvalues and eigenstates, we need to choose a basis. The natural choice corresponds to arrays of up and down spins
in the $z$ direction, as in Eqs.~(\ref{flipflop}), (\ref{upup}) and (\ref{updown}). We refer to it as the site basis. In this basis, $H_z$ and the
Ising interaction contribute to the diagonal elements of the matrix, and the flip-flop
term leads to the off-diagonal elements. 

In the absence of the Ising interaction, the excitations move freely through the chain.
In this case the eigenvalues and eigenstates can be found analytically. 
The existence of an analytical method to 
find the spectrum of a system guarantees its integrability.
The addition of the Ising interaction eventually leads to the onset of quantum chaos.
The source of chaos is the interplay between the Ising interaction and 
the defect.\cite{Santos2004} 

To bring the system to the chaotic regime, we set $J_{xy}=1$ (arbitrary units), 
choose $J_z=\epsilon_d=0.5$ (arbitrary units), and place
the defect on site $d=\left\lfloor L/2 \right\rfloor$, where $\lfloor L/2 \rfloor$ stands for the largest integer less than or equal to $L/2$. These choices are based
on the following factors.
(a) The strength of the Ising interaction cannot be
much larger than $J_{xy}$, because if it were, basis vectors with different numbers of pairs 
of parallel spins would have very different energies and $J_{xy}$ would not be able to effectively couple them. For example, the energy difference between $|\uparrow \uparrow \downarrow \rangle$
and $|\uparrow \downarrow \uparrow \rangle$ is $J_z/2$; if we had $J_z \gg J_{xy}$, the matrix element $J_{xy}/2$ coupling these two basis vectors would become ineffective. As a result, the eigenstates would involve only a small portion of the
basis vectors, which would mean localized eigenstates and therefore non-chaotic systems.
(b) The defect cannot be placed on the edges of the chain, because it has been shown~\cite{Alcaraz1987} that in this case an analytical solution exists and the system is therefore
still integrable. (c) We cannot have $\epsilon_d \gg J_{xy}$, because this would break the chain in two; that is, an excitation on one side
of the chain would not have enough energy to overcome
the defect and reach the other side of the chain.
In effect, we would be dealing with two independent chains described by the XXZ model,
which is integrable.\cite{BetheIntro,Karbach}

\section{Quantum chaos}

We use the level spacing distribution to identify
when the system becomes chaotic. We analyze also what happens to the 
structure of the eigenstates once level repulsion occurs. 

\subsection{Level spacing distribution}

The distribution $P(s)$ of the spacings, $s$, of neighboring energy levels differs depending on the regime of the system.\cite{HaakeBook,Guhr1998,ReichlBook} 
The energy levels of 
integrable systems are not correlated, and are not prohibited from crossing, 
so the distribution is Poissonian (P), 
\begin{equation}
P_{\rm P}(s) = e^{-s}.
\end{equation} 
In chaotic systems the eigenvalues become correlated and crossings are avoided. There is level repulsion, 
and $P(s)$ is given by the Wigner-Dyson (WD) distribution, as
predicted by random matrix theory. 
The form of the Wigner-Dyson distribution depends on the symmetry properties 
of the Hamiltonian. 
Systems with time reversal invariance are described by a
Gaussian orthogonal ensemble, which corresponds to an ensemble of real symmetric matrices, whose elements $H_{ij}$ are
independent random numbers chosen from a Gaussian distribution.
The average of the elements and the variance satisfy $\langle H_{ij}\rangle=0$ and
$\langle H_{ij}^2\rangle=1+\delta_{ij}$.
The level spacing distribution of a Gaussian orthogonal ensemble is given by
\begin{equation} 
P_{\rm WD}(s) = \frac{\pi s}{2}e^{-\pi s^2/4}.
\end{equation}

The Hamiltonian in Eqs.~(\ref{hamall}) is also real and symmetric,
so the distribution achieved in the chaotic limit is the same as $P_{\rm WD}(s)$. 
However, our system has only short-range-two-body interactions, so it cannot reproduce all features of the Gaussian orthogonal ensemble. The density of states, 
for instance, is Gaussian for Eqs.~(\ref{hamall}),
but for the Gaussian orthogonal ensemble it has the shape of half of a circle. Interested readers can obtain these forms using the codes provided.\cite{www} Another difference is the
structure of the eigenstates, as discussed in Sec.~\ref{princomp}. 
For Gaussian orthogonal ensembles, all the eigenstates are statistically similar and highly delocalized, whereas
for the Hamiltonian in Eqs.~(\ref{hamall}), delocalization is restricted to the middle of the spectrum.

To obtain the level spacing distribution, we first need to separate the eigenvalues
according to their symmetry sectors (subspaces). 
If we mix eigenvalues from different symmetry sectors, we cannot achieve a Wigner-Dyson distribution even if the system is chaotic 
because eigenvalues from different subspaces are independent and therefore uncorrelated,
so do not repel each other. We discuss further the danger of mixing
eigenvalues from different symmetry sectors in Sec.~\ref{symm}. For the moment,
we need to remember only that the Hamiltonian in Eqs.~(\ref{hamall}) conserves $S^z$. This symmetry
implies that the Hamiltonian 
matrix is separated into uncoupled blocks, corresponding to subspaces with a fixed number of spins 
in the up direction. In the following, we select a particular subspace with $L/3$ up spins, whose Hamiltonian matrix has therefore dimension $\dimm =L!/[(L/3)!(L-L/3)!]$. 

The second essential step
before computing the spacing distribution is to unfold the spectrum.
The procedure consists of locally rescaling the energies,
so that the local density of states of the renormalized eigenvalues is 1.
This rescaling allows for the comparison of spectra obtained for different parameters
and for different systems. There are different ways to unfold the spectrum. 
A simple and commonly used procedure is to order the spectrum in increasing values of energy;
separate it into several smaller sets of eigenvalues; and divide each eigenvalue by the mean level spacing of its particular set.
The mean level spacing of the new set of renormalized energies becomes 1.
Because the density of states is the number of states in an interval of energy, that 
is, the reciprocal of the mean level spacing, this procedure also ensures that
the local density of states is unity. This procedure is the one we used.\cite{www}

Given the unfolded spacings of neighboring levels, the histogram can now be computed. 
To compare it with 
the theoretical curves, the distribution needs to be normalized, so that its total area 
is equal to 1.

\begin{figure}[htb]
\includegraphics[width=0.4 \textwidth]{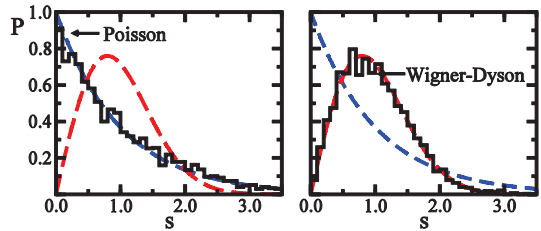}
\caption{(Color online) Level spacing distribution for the Hamiltonian in Eqs.~(\ref{hamall}) with 
$L=15$, 5 spins up, $\omega=0$, 
$\epsilon_d=0.5$, $J_{xy}=1$, and $J_z=0.5$ (arbitrary units); bin size = 0.1. 
(a) Defect on site $d=1$;(b) defect 
on site $d=7$. The dashed lines are the theoretical curves.}
\label{fig:Ps}
\end{figure}

Figure~\ref{fig:Ps} shows the level spacing distribution when the defect is 
placed on site 1 and on site $\left\lfloor L/2 \right\rfloor$. The first case
corresponds to an integrable model and the distribution is a Poisson; the second case is 
a chaotic system, so the distribution is Wigner-Dyson. 

\subsection{\label{princomp}Number of principal components}

We now investigate how the transition from a Poisson to a Wigner-Dyson
distribution affects the structure of the eigenstates. In particular, we 
study how delocalized they are in both regimes.

To determine the spreading of the eigenstates in a particular
basis, we look at their components. Consider an eigenstate $|\psi_i \rangle $ written in the basis vectors $|\xi_k\rangle$ as 
$|\psi_i \rangle= \sum_{k=1}^{\dimm} c_{ik} |\xi_k\rangle$. It
will be localized if it has 
the participation of few basis vectors, that is, if a few $|c_{ik}|^2$ 
make significant contributions. It will be delocalized if
many $|c_{ik}|^2$ participate with similar values.
To quantify this criterion, we use the 
sum of the square of the probabilities, $|c_{ik}|^4$ (the sum of
the probabilities would not be a good choice, because normalization implies
$\sum_{k=1}^{\dimm} |c_{ik}|^2=1$), and define the
number
of principal components of eigenstate $i$ as\cite{Izrailev1990,ZelevinskyRep1996}
\begin{equation}
n_i \equiv \frac{1}{\sum_{k=1}^{\dimm} |c_{ik}|^4}.
\label{NPC}
\end{equation}
The number
of principal components gives the number of basis vectors which contribute to each eigenstate. It is 
small when the state is localized and large when the state is delocalized.

For Gaussian orthogonal ensembles, the eigenstates are random vectors, that is, the amplitudes $c_{ik}$ 
are independent random variables. These states are completely delocalized. 
Complete delocalization does not mean,
however, that the number
of principal components is equal to $\dimm$. Because the weights $|c_{ik}|^2$ fluctuate, the average over the ensemble 
gives $\mbox{number
of principal components} \sim \dimm/3$.\cite{Izrailev1990,ZelevinskyRep1996}

To study the number
of principal components for Eqs.~(\ref{hamall}), we need to choose a basis.
This choice depends on the question we want to address.
We consider two bases, the site- and mean-field basis.
The site-basis is appropriate when analyzing the spatial delocalization of the system.
To separate regular from chaotic behavior, a more appropriate basis consists of 
the eigenstates of the integrable limit of the model, which is known as 
the mean-field basis.\cite{ZelevinskyRep1996} 
In our case the integrable limit corresponds to Eqs.~(\ref{hamall}) with
$J_{xy} \neq 0$, $\epsilon_d\neq 0$, and $J_z=0$. 

We start by writing the Hamiltonian in the site-basis. Let us denote these 
basis vectors by $|\phi_j \rangle$. In the absence of the Ising interaction, the diagonalization of the 
Hamiltonian leads to the mean-field basis vectors.
They are given by $|\xi_k \rangle = \sum_{j=1}^{\dimm} b_{kj} |\phi_j \rangle$.
The diagonalization of the complete matrix, including the Ising interaction, gives
the eigenstates in the site-basis, $|\psi_i \rangle = \sum_{j=1}^{\dimm} a_{ij} |\phi_j \rangle$.
If we use the relation between $|\phi_j \rangle$ and $|\xi_k \rangle$, we may 
also write the eigenstates of the total Hamiltonian in Eqs.~(\ref{hamall}) in the mean-field basis as
\begin{equation}
|\psi_i \rangle = \sum_{k=1}^{\dimm} \left( \sum_{j=1}^{\dimm} a_{ij} b^{*}_{kj} \right) 
|\xi_k \rangle= \sum_{k=1}^{\dimm} c_{ik} |\xi_k \rangle.
\end{equation}

\begin{figure}[htb]
\includegraphics[width=0.4\textwidth]{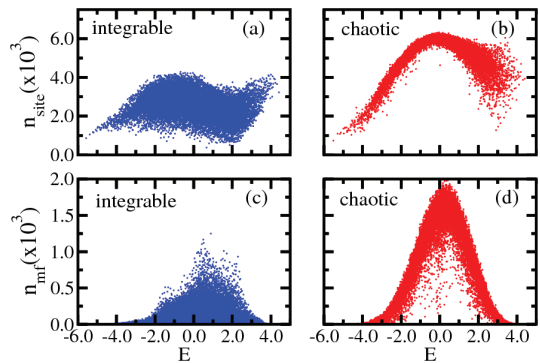}
\caption{(Color online) Number of principal components for the eigenstates
of the Hamiltonian in Eqs.~(\ref{hamall}) versus energy; 
$L=18$, 6 spins up, $\omega=0$, 
$\epsilon_d=0.5$, $J_{xy}=1$, and $J_z=0.5$ (arbitrary units). (a) and (b)
site-basis; (c) and (d) mean-field basis;
(a) and (c) defect on site $d=1$; (b) and (d) defect 
on site $d=9$.}
\label{fig:NPC}
\end{figure}

Figures~\ref{fig:NPC} shows the number
of principal components for the eigenstates in the site-basis [(a), (b)] and in the mean-field
basis [(c), (d)] for the cases where the defect is 
placed on site 1 [(a), (c)] and on site $\left\lfloor L/2 \right\rfloor$ [(b), (d)]. 
The level of delocalization increases significantly in the chaotic regime.
However, contrary to random matrices, the largest values are
restricted to the middle of the spectrum, the states at the edges
being more localized. This property is a consequence of the Gaussian shape of the density of states of 
systems with two-body interactions.
The highest concentration of states appears 
in the middle of the spectrum, where the
strong mixing of states can occur leading to widely distributed eigenstates.

An interesting difference between the integrable and chaotic regimes is
the fluctuations of the number
of principal components. For the regular system the
number
of principal components shows large fluctuations. In contrast, in the chaotic regime the
number
of principal components approaches a smooth function of energy. 
Chaotic eigenstates close in energy have similar structures
and consequently similar values of the number
of principal components.

\section{Symmetries\label{symm}}

The presence of a defect breaks symmetries of the system. In this section we remove the defect
and have a closer look at the symmetries. 

We refer to the system in the absence of a defect ($\epsilon_d =0$) as defect-free.
Contrary to the case where $\epsilon_d \ne 0$,
a defect-free spin-1/2 chain with NN couplings remains integrable even when the Ising 
interaction is added. This system can be analytically solved using the Bethe ansatz.\cite{BetheIntro,Karbach} To drive the system to chaos, 
while keeping it defect free, we need to add further couplings.
By considering couplings between next-nearest-neighbors 
(NNNs),\cite{Hsu1993,Kudo2004,Santos2009,Santos2010PRE} 
the Hamiltonian becomes
\begin{equation}
H = H_{\text{NN}} + \alpha H_{\text{NNN}},
\label{hamNNN} 
\end{equation}
where
\begin{equation}
H_{\text{NNN}} = \sum_{n=1}^{L-2} \left[ J_{xy}' \left(
S_n^x S_{n+2}^x + S_n^y S_{n+2}^y \right) +
J_z' S_n^z S_{n+2}^z \right].
\end{equation}
For sufficiently large $\alpha$ ($\alpha \gtrsim 0.2$ for $L=15$), there are various
scenarios for which chaos can develop, which include
the absence of Ising interactions, $J_z=J_z'=0$;
the absence of the flip-flop term between next-nearest-neighbors, $J_{xy}'=0$;
the absence of Ising interaction between next-nearest-neighbors, $J_{z}'=0$;
and the presence of all four terms.

Depending on the parameters in Eq.~(\ref{hamNNN}), we
might not obtain a Wigner-Dyson distribution even if the system is chaotic if not all symmetries of the system are taken into account.\cite{Kudo2004,Santos2009}
We have mentioned conservation of total spin in the $z$ direction.
In the absence of a defect other symmetries of $H$ in Eq.~(\ref{hamNNN}) include the following.\cite{Brown2008}

{\em Parity}. Parity may be understood by imagining
a mirror in one edge of the chain. 
For eigenstates written in the site-basis,
the probability of each basis vector is equal to that of its reflection.
For example, suppose we have $L=4$ and one excitation. The eigenstates are
given by $|\psi_i \rangle = a_{i1}|\uparrow \downarrow \downarrow \downarrow \rangle +
a_{i2} |\downarrow \uparrow \downarrow \downarrow \rangle +
a_{i3} |\downarrow \downarrow \uparrow \downarrow \rangle +
a_{i4} |\downarrow \downarrow \downarrow \uparrow \rangle$. 
The amplitudes are either $a_{i1}=a_{i4}$ and $a_{i2}=a_{i3} $ for even 
parity or $a_{i1}=- a_{i4}$ and $a_{i2}=- a_{i3} $ for odd 
parity. The level spacing distribution needs to be independently obtained for 
each parity.

{\em Spin reversal}. If the chain has an even number of sites and $L/2$ up spins,
then $S^z=0$. In this sector 
pairs of equivalent basis vectors correspond to those which become equal if we rotate all the
spins from one vector by 180$^\circ$. For example, state $|\uparrow \downarrow \downarrow \uparrow\rangle$
pairs with state $|\downarrow \uparrow \uparrow \downarrow \rangle$.

{\em Total spin}. If the system is isotropic, that is, $J_{xy}=J_z$ and $J_{xy}'=J_z'$, the
total spin, $S^2=(\sum_{n=1}^L \vec{S}_n)^2$, is conserved.

For closed boundary conditions there is also momentum conservation. The more
symmetries the system has, the smaller the subspaces become for a given system size, 
which is not good for statistics. For this reason 
we chose open boundary conditions.

\begin{figure}[htb]
\includegraphics[width=0.37\textwidth]{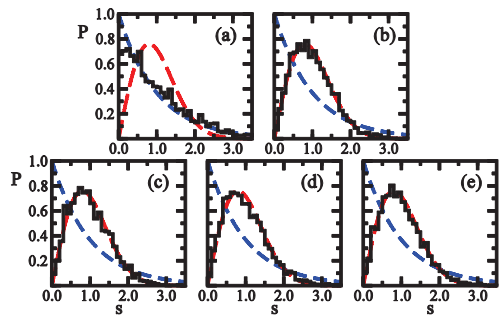}
\caption{(Color online) Level spacing distribution for
Eq.~(\ref{hamNNN}) with $\alpha=0.5$, $J_{xy}=1$; 
$\mbox{bin size} = 0.1$.
(a) $L=14$, 7 spins up, $J_{xy}'=J_{z}=J_{z}'=1$. All eigenvalues of the subspace $S^z=0$ are considered. (b) $J_{xy}'=1$, $J_{z}=J_{z}'=0.5$.
(c) $J_{xy}'=1$, $J_{z}=J_{z}'=0$. 
(d) $J_{xy}'=0$, $J_{z}=J_{z}'=0.5$.
(e) $J_{xy}'=1$, $J_{z}=0.5$, $J_{z}'=0$.
For (b)--(e) $L=15$, 5 spins up. 
The eigenvalues are separated according to 
the parity of the corresponding eigenstates.
$P(s)$ is the average of the distributions 
of the two parity sectors.}
\label{fig:symmetry}
\end{figure}

In Fig.~\ref{fig:symmetry} we show the level spacing distribution for the four chaotic 
systems we have described. All the figures involve eigenvalues of 
a single selected $S^z$-sector. Both Figs.~\ref{fig:symmetry}(a) and \ref{fig:symmetry}(b) show results for a chaotic spin-1/2 chain with all interactions in Eq.~\eqref{hamNNN} included. Because Fig.~\ref{fig:symmetry}(a) mixes the three symmetries we have discussed,
$P(s)$ becomes a Poisson distribution, even though the system is in fact chaotic. In Fig.~\ref{fig:symmetry}(b), just as in Figs.~\ref{fig:symmetry}(c)-(e), we 
circumvent the $S^z=0$ subspace by avoiding chains with an even number of sites. 
We also choose $J_z \neq J_{xy}$ to avoid conservation of total spin.
By doing so, the only remaining symmetry is parity, which we take into account.
The expected Wigner-Dyson distributions are then obtained.

\section{Discussion}

The computer programs used to obtain the data for Figs.~\ref{fig:Ps}, \ref{fig:NPC}, and
\ref{fig:symmetry} are available.\cite{www} The reader will also find
programs to study Gaussian orthogonal ensembles and suggestions for other investigations.

Spin 1/2 chains are excellent models for introducing students to some of the basic concepts
of linear algebra, quantum mechanics, as well as to current areas of research. In addition to the crossover from integrability to chaos, they can be
used to introduce topics as diverse as 
the metal-insulator transition, quantum phase transition, entanglement~\cite{Amico2008}, 
spintronics, and methods of quantum control~\cite{Vandersypen2005}. 
They have been considered as models for quantum computers~\cite{Kane1998} and magnetic compounds~\cite{Sologubenko2001}, and recently have been
simulated in optical lattices.\cite{Trotzky2008,Simon2011,ChenARXIV} 

\begin{acknowledgments}
A.~G.\ thanks Stern College for Women of Yeshiva University for a summer fellowship and 
the Kressel Research Scholarship for a one-year financial support.
This work is part of her thesis for the S.\ Daniel Abraham Honors Program. 
L.~F.~S. thanks M. Edelman, F.\ M.\ Izrailev, A. Small, and F.\ Zypman for useful discussions.
\end{acknowledgments}

\end{document}